\documentclass[12pt,showpacs,superscripta ddress,preprintnumbers,amsmath,amssymb]{revtex4}
\usepackage{epsfig} 
\usepackage{amsmath} 
\usepackage{graphicx} 
\usepackage{subfigure}
\large

\def\gev{{\rm GeV}}

\def\bsll{B_s \rightarrow \mu^+ \, \mu^-} 
\def\bkll{B \rightarrow K \mu^+ \, \mu^-} 
\def\afb{\left\langle A_{FB}\right\rangle}
\def\dfb{A_{FB}(z)}
\def\incl{B \rightarrow X_s \mu^+ \, \mu^-}

\def\lesssim{\mathrel{\hbox{\rlap{\hbox{\lower4pt\hbox{$\sim$}}}\hbox{$<$}}}} 
\def\gtrsim{\mathrel{\hbox{\rlap{\hbox{\lower4pt\hbox{$\sim$}}}\hbox{$>$}}}}

\newcommand{\spp}{\vphantom{\bigg(}}

\begin{document} 
\title{\bf Large forward-backward asymmetry in $B \to K \mu^+ \mu^-$ \\
from new physics tensor operators}
 
\author{Ashutosh Kumar Alok}
\email{alok@theory.tifr.res.in}
\affiliation{Tata Institute of Fundamental Research, Homi Bhabha
Road, Mumbai 400005, India}

\author{Amol Dighe}
\email{amol@theory.tifr.res.in}
\affiliation{Tata Institute of Fundamental Research, Homi Bhabha
Road, Mumbai 400005, India}
                                                           
\author{S. Uma Sankar}
\email{uma@phy.iitb.ac.in}
\affiliation{Indian Institute of Technology Bombay, Mumbai 400076,
India}

\date{\today} 

\preprint{TIFR/TH/08-41}


\begin{abstract}
We study the constraints on possible new physics contribution 
to the forward-backward asymmetry of muons, $A_{FB}(q^2)$,  
in $B \rightarrow K \mu^+ \mu^-$.
New physics in the form of vector/axial-vector operators 
does not contribute to $A_{FB}(q^2)$ whereas new physics
in the form of scalar/pseudoscalar operators can enhance 
$A_{FB}(q^2)$ only by a few per cent. 
However new physics the form of tensor operators 
can take the peak value of $A_{FB}(q^2)$ to as high as $40\%$
near the high-$q^2$ end point.
In addition, if both scalar/pseudoscalar and tensor 
operators are present, then $A_{FB}(q^2)$ can be more than
$15\%$ for the entire high-$q^2$ region $q^2 > 15$ GeV$^2$. 
The observation of significant $A_{FB}$ would imply the
presence of new physics tensor operators, whereas its
$q^2$-dependence could further indicate the presence of new 
scalar/pseudoscalar physics.
\end{abstract}

\maketitle 

\newpage

\section{Introduction}

Flavor changing neutral interactions (FCNI) are forbidden at the
tree level in the standard model (SM). Therefore they have the
potential to test higher order corrections to the SM and also
constrain many of its possible extensions. Among all FCNI, rare $B$
decays play an important role in searching new physics beyond
the SM. The quark level FCNI $b \to s \mu^+ \mu^-$ is
responsible for (i) the inclusive semileptonic decay $B \to X_s \mu^+ \mu^-$, 
(ii) the exclusive semileptonic decays $B \to
(K, K^*) \mu^+ \mu^-$, and (iii) the purely leptonic decay $B_s
\rightarrow \mu^+ \mu^-$. 
Both the inclusive and exclusive semileptonic decays
have been observed experimentally 
\cite{babar04_incl,belle05_incl,babar-03, babar-06, belle-03,hfag}
with branching ratios close to their SM predictions 
\cite{ali-02,lunghi,kruger-01,isidori}.

In \cite{alok-sankar01}, the impact of these
measurement on the new physics contribution to the branching ratio
 $B(\bsll)$ was considered. It was shown that new physics
in the form of vector/axial-vector operators is severely constrained
by the data on  $B(B \rightarrow K \mu^+ \mu^-)$ and 
$B(B \rightarrow K^* \mu^+ \mu^-)$, 
so an order of magnitude enhancement in the branching ratio
of $B_{s}\rightarrow\mu^{+}\mu^{-}$ is ruled out. On the other hand,
if new physics is in the form of scalar/pseudoscalar operators, then
$B(B \to K^* \mu^+ \mu^-)$ does not put any useful constraint on the new
physics couplings and allows an order of magnitude enhancement in
 the $B(\bsll)$. Therefore 
$B(B_s \rightarrow \mu^+ \mu^-)$ is sensitive
to an extended Higgs sector. In \cite{tension}, the 
constraints on scalar/pseudoscalar new physics contribution to the
$B(B \rightarrow K \mu^+ \mu^-)$ were studied. 
It was shown that a large deviation in 
$B(\bkll)$ from its SM prediction is not possible.

In \cite{ali-92}, the forward-backward (FB) asymmetry of leptons  
in semileptonic decays of
mesons was introduced as an observable sensitive to the physics beyond the
SM. In particular, the FB asymmetry of muons,
$A_{FB}$, in $\bkll$ is important because its value is negligibly small in the
SM \cite{ali-00}. This is due to the fact that hadronic current for 
$B \rightarrow K$
transition does not have any axial vector contribution;
it can have a nonzero value only if it receives contribution from new
physics. The sensitivity of 
$A_{FB}$ for testing non-standard Higgs sector
has been studied in literature in detail
\cite{yan-00,bobeth-01,erkol-02,demir-02,li-04}. However in
\cite{alok-02}, it was shown that the present upper bound on the branching ratio
of $B_s \rightarrow \mu^+ \mu^-$ \cite{cdf-07} restricts 
the average (or integrated) FB asymmetry, $\afb$, to about $1\%$ as long as 
the only new physics is in the form of scalar/pseudoscalar operators. 
Such a small FB asymmetry is very difficult to be measured 
in experiments and hence searching for new  
scalar/pseudoscalar physics through $\afb$ will be a futile exercise.

The forward-backward asymmetry can also get contributions from
tensor operators.
In the SM, the tensor operators in $b\to s\mu^+ \mu^-$ 
arise at higher order in the electroweak operator product expansion 
from finite external momenta in the matching calculations, 
however their contribution is negligibly small and we shall not
consider them in this paper.
However in models beyond the SM, tensor operators may contribute
significantly to the decay and to the asymmetry $A_{FB}$.
For example, in the minimal supersymmetric standard model (MSSM), 
the tensor operators arise from photino and zino box diagrams at
the leading order operator product expansion \cite{Bobeth:2007dw}. 
Tensor operators can also be induced by scalar operators under 
renormalization group running \cite{Hiller:2003js,Borzumati:1999qt}. 
In leptoquark models, tensor operators are induced by the interactions 
of leptoquarks with the SM Higgs field \cite{Hirsch:1996qy}.

In \cite{Bobeth:2007dw}, the effect of these
operators to $\afb$ was studied,
where it was shown that $\afb$ can be as high as $3\%$ at $90\%$ C.L.  
if new physics is only in the form of tensor operators, 
whereas it can rise to $15\%$ if both scalar/pseudoscalar
and tensor  new physics operators are present. 
The integrated asymmetry $\afb$ has been measured by BaBar
\cite{babar-06} and Belle \cite{belle-06,ikado-06} to be 
\begin{equation} 
\left\langle A_{FB}\right\rangle  =  (0.15_{-0.23}^{+0.21} \pm 0.08)\, 
\, \, \, \, \, 
 {\rm (BaBar)} \, , 
\end{equation} 
\begin{equation} 
\left\langle A_{FB}\right\rangle  = (0.10 \pm 0.14 \pm 0.01) \, \, 
\, \, {\rm (Belle)}. \label{fb_exp} 
\end{equation} 
These measurements are consistent with zero. 
However, they can be as high as $\sim 40\%$ within $2\sigma$ error bars.
Future experiments like a Super-$B$ factory or the LHC will increase 
the statistics by more than two orders of magnitude.
For example at ATLAS, the number of expected $\bkll$
events even after analysis cuts is expected to be $\sim 4000$ with 
$30$ fb$^{-1}$ data \cite{Adorisio},
which will be collected within the first three years.
Thus, $\langle A_{FB} \rangle$ can soon be probed to values as low as $5\%$. 

With higher statistics, one will be able to determine even  
the distribution of $A_{FB}$ as a function of the invariant dilepton
mass squared $q^2$, which can provide a stronger handle on 
this quantity than just its
average value $\langle A_{FB} \rangle$.
Moreover, since the theoretical predictions for the rate of
$\bkll$ are rather uncertain in the intermediate $q^2$ region
($7$~GeV$^2 < q^2 < 12$~GeV$^2$) owing to the vicinity of
charmed resonances, it is important to look at
the quantity $A_{FB}(q^2)$ in the complete $q^2$ range so that
its robust features may be identified. Indeed, it turns out 
that with the new physics considered in this paper,
$A_{FB}(q^2)$ is high near the high-$q^2$ end point.

In this paper we study $A_{FB}(q^2)$ in the complete $q^2$ region and 
explore the possibility of large FB asymmetry in some specific regions of the 
dilepton invariant mass spectrum. 
This paper is organized as follows. In section \ref{fbasy}, 
we present the theoretical expressions 
for the FB asymmetry of $\bkll$ considering new physics 
in the form of scalar/pseudoscalar and 
tensor operators. 
In section \ref{spnp} we study $A_{FB}(q^2)$ due to new physics only in the 
form of scalar/pseudoscalar operators whereas in section 
\ref{tpnp} we consider $A_{FB}(q^2)$ due to new physics only
in the form of tensor operators. 
In section \ref{sptp}, we calculate $A_{FB}(q^2)$
when both the scalar/pseudoscalar we well as tensor
operators are present.
Finally in section \ref{concl}, we present 
the conclusions.

\section{Forward-backward asymmetry of muons in $\bkll$ }
\label{fbasy}

We consider new physics in the form of scalar/pseudoscalar and tensor operators. 
The effective Lagrangian for the quark level transition $b \rightarrow s
\mu^+ \mu^-$ can be written as
\begin{equation}
L(b \rightarrow s \mu^{+} \mu^{-}) = L_{SM} + L_{SP} + L_{T} \;,
\label{ltotal}
\end{equation}
where
\begin{eqnarray}
L_{SM} &=&  \frac{\alpha G_F}{\sqrt{2} \pi} V_{tb} V^\star_{ts}
\biggl\{ C^{\rm eff}_9        (\bar{s} \gamma_\mu P_L b)    \,
\bar{\mu} \gamma_\mu \mu  +
C_{10}              (\bar{s} \gamma_\mu P_L b)        \,   \bar{\mu} \gamma_\mu \gamma_5 \mu \nonumber \\
&&-
2 \frac{C^{\rm eff}_7}{q^2} m_b \, (\bar{s} i \sigma_{\mu\nu} q^\nu P_R b) \, \bar{\mu} \gamma_\mu \mu
\biggr\}\;, 
\label{sml}\\
L_{SP} &=& \frac{\alpha G_F}{\sqrt{2} \pi} V_{tb} V^\star_{ts}
\biggl\{R_S ~\bar{s} P_R b ~\bar{\mu}\mu + 
 R_P ~\bar{s} P_R b \, \bar{\mu}\gamma_5 \mu
 \biggr\} \;, 
\label{sp}\\
 L_{T} &=& \frac{\alpha G_F}{\sqrt{2} \pi} V_{tb} V^\star_{ts}
\biggl\{C_T ~\bar{s} \sigma_{\mu \nu } b ~\bar{\mu}
\sigma^{\mu\nu}\mu + i C_{TE}  ~\bar{s} \sigma_{\mu \nu } b
~\bar{\mu} \sigma_{\alpha \beta } \mu ~\epsilon^{\mu \nu \alpha
\beta} \biggr\} \;.
\label{ten}
\end{eqnarray}
Here $P_{L,R} = (1 \mp \gamma_5)/2$ and $q_{\mu}$ is the sum of 4-momenta of 
$\mu^+$ and $\mu^-$. $R_S$ and $R_P$ are new physics scalar/pseudoscalar 
couplings whereas $C_T$ and $C_{TE}$ are new physics tensor
couplings. 

Within the SM, the Wilson coefficients in eq.~(\ref{sml})  
have the following values: 
\begin{equation} 
C_{7}^{\rm eff} = -0.310 \; , \;  
C_{9}^{\rm eff} = +4.138 + Y(q^2) \; , \; C_{10} = -4.221\;, 
\end{equation} 
where the function $Y(q^2)$ is given by \cite{buras-95,misiak-95}
\begin{eqnarray}
Y(q^2)&=& g(m_c,q^2)(3 C_1 + C_ 2 +3C_3 + C_4 + 3C_5 + C_6)
-\frac{1}{2} g(0,q^2)(C_3 + 3 C_4)\nonumber\\
&-&\frac{1}{2}g(m_b,q^2)(4C_3 + 4C_4 +3C_5+C_6)
+\frac{2}{9}(3C_3 +C_4+3c_5+C_6)\; .
\label{y-q2}
\end{eqnarray}
Here we take the values of the relevant Wilson coefficients to be
\begin{eqnarray}
C_1 = -0.249,\ C_2 = 1.107,\ C_3 = 0.011,\nonumber \\
 C_4 = -0.025, \ C_5 = 0.007, \ C_6 = -0.031,
\end{eqnarray}
all of which are computed at the scale $\mu = m_b = 5$ GeV.
The function $g$ is given by
\begin{eqnarray}
\lefteqn{g(m_i,q^2)=-\frac{8}{9}\ln(m_i/m_b^{\text{pole}})+\frac{8}{27}+\frac{4}{9}y_i
-\frac{2}{9}(2+y_i)\sqrt{|1-y_i|}}\nonumber\\[.7ex]
&&\times\left\{
\Theta(1-y_i)\left[\ln\left(\frac{1 + \sqrt{1-y_i}}{1 - \sqrt{1-y_i}}\right)-i\pi \right]+ \Theta(y_i-1) \ 2 \tan^{-1} \left(\frac{1}{\sqrt{y_i-1}} \right) \right\},
\end{eqnarray}
with $y_i \equiv 4m_i^2/q^2$.

The normalized FB asymmetry is defined as 
\begin{eqnarray} 
A_{FB}(z)= \frac{\int_0^{1}dcos\theta \frac{d^2\Gamma}{dz \ 
dcos\theta}-\int_{-1}^{0}dcos\theta \frac{d^2\Gamma}{dz\ dcos\theta}} 
{\int_0^{1}dcos\theta \frac{d^2\Gamma}{dz\ dcos\theta}+\int_{-1}^{0}dcos\theta \frac{d^2\Gamma}{dz 
\ d\cos\theta}}\;. 
\end{eqnarray} 
with $z \equiv q^2/m_{B}^{2}$.
In order to calculate the FB asymmetry, we first need to 
calculate the differential decay width. 
The decay amplitude for $B(p_1)\to K(p_2)\,\mu^+(p_+)\,\mu^-(p_-)$ is given by 
\begin{eqnarray} 
M\,(B\rightarrow K \mu^{+}\mu^{-}) &=& \frac{\alpha G_F}{2\sqrt{2} \pi} V_{tb} V^\star_{ts}   
\nonumber \\ 
&\times& 
\Bigg[\left< K(p_2) \left|\bar{s}\gamma_{\mu}b\right|B(p_1)\right> \left\{C_{9}^{\rm eff}\bar{u}(p_-)\gamma_{\mu}v(p_+)  
+C_{10}\bar{u}(p_-)\gamma_{\mu}\gamma_{5} v(p_+)\right\}  
\nonumber \\ 
& & - 2\frac{C^{\rm eff}_7}{q^2} m_b \left< K(p_2)\left|\bar{s}i\sigma_{\mu\nu}q^{\nu}b\right|B(p_1)\right> \;
\bar{u}(p_-)\gamma_{\mu}v(p_+) 
\nonumber \\ 
& & +\left< K(p_2)\left|\bar{s}b\right|B(p_1)\right>\;\left\{R_S \bar{u}(p_-)v(p_+)+
R_P\bar{u}(p_-)\gamma_5 v(p_+)\right\} \nonumber\\
& & +2C_T \left< K(p_2)\left|\bar{s}\sigma_{\mu\nu}b\right|B(p_1)\right>\;\bar{u}(p_-)\sigma^{\mu\nu}v(p_+)
\nonumber \\ 
& & + 2iC_{TE}\epsilon^{\mu \nu \alpha\beta}\left< K(p_2)\left|\bar{s}\sigma_{\mu\nu}b\right|B(p_1)\right>\;
\bar{u}(p_-)\sigma_{\alpha \beta}v(p_+) \Bigg]\; ,
\label{matrix}
\end{eqnarray} 
where $q_\mu = (p_1-p_2)_\mu = (p_+ + p_-)_\mu$.
The relevant matrix elements are
\begin{eqnarray} 
\left< K(p_2) \left|\bar{s}\gamma_{\mu}b\right|B(p_1)\right> &=& 
(2p_1-q)_{\mu}f_{+}(z)+(\frac{1-k^2}{z})\, q_{\mu}[f_{0}(z)-f_{+}(z)]\;, \\
\left< K(p_1)\left|\bar{s}i\sigma_{\mu\nu}q^{\nu}b\right|B(p_1)\right> &=& \Big[(2p_1-q)_{\mu}q^2-(m_{B}^{2}-m_{K}^{2})q_{\mu}\Big]\,\frac{f_{T}(z)}{m_B+m_{K}}\;, \\
\left< K(p_2)\left|\bar{s}b\right|B(p_1)\right> &=& \frac{m_B(1-k^2)}{\hat{m}_b}f_0(z)\;, \\
\left< K(p_2)\left|\bar{s}\sigma_{\mu\nu}b\right|B(p_1)\right> 
&=& -i\Big[(2p_1-q)_{\mu}q_{\nu}-(2p_1-q)_{\nu}q_{\mu}\Big]\,\frac{f_T}{m_B+m_{K}}\; ,
\end{eqnarray}
where $k \equiv m_K/m_B$  and $\hat{m}_b \equiv m_b/m_B$.

Using the above matrix elements, the double differential decay widths can be calculated as
\begin{eqnarray}
\frac{d^{2}\Gamma}{dzdcos\theta} & = & \frac{G_{F}^{2}\alpha^{2}}{2^{11}\pi^{5}}\,|V_{tb}V_{ts}^{*}|^{2}\, m_{B}^{5}\,\phi^{1/2}  \nonumber \\ 
&\times&
\Bigg[ z\left\{\frac{\hat{m}_\mu}{m_B}{\rm Re} (CE^*)+\frac{1}{4m_B^{2}}(|E|^2+\beta_{\mu}^2|D|^2)\right\}
\nonumber \\
&&+\phi \left\{ \frac{1}{4} (|A|^2+|B|^2)\,+\,2\hat{m}_\mu \, m_B\, {\rm Re}(AF^*) \right\}
\nonumber \\
&&  + (1-k^2)\left\{2\hat{m}_\mu^2 {\rm Re}(BC^*)+\frac{\hat{m}_\mu}{m_B}{\rm Re}(BE^*)\right\}
\nonumber \\
&&  + \hat{m}_\mu^2 \left\{ (2+2k^2-z)|B|^2 + z|C|^2 \right\}+\phi\, z\, m_B^2\,(1-\beta_{\mu}^2)|F|^2
\nonumber \\
&&  + \phi \beta_{\mu}^2 \left\{z\, m_B^2 (|F|^2 + 4|G|^2)-\frac{1}{4}(|A|^2+|B|^2) \right\} \cos^2\theta
\nonumber \\
&& - \Bigg. \phi^{1/2}\beta_{\mu}  \left\{\frac{\hat{m}_\mu}{m_B}{\rm Re} (AD^*)+4m_\mu(1-k^2){\rm Re} (BG^*) 
 +4z\hat{m}_\mu m_B {\rm Re}(CG^*) \nonumber \right. \Bigg. \\
&& + \left.  \Bigg. 2z{\rm Re}(GE^*)+\frac{z}{4}{\rm Re} (DF^*)\right\}  \Bigg. \cos\theta \Bigg]\;,
\label{double_drate}
\end{eqnarray}
where 
\begin{eqnarray}
\hat{m}_\mu & \equiv & m_\mu/m_B \nonumber \\
\phi & \equiv  & 1+k^{4}+z^{2}-2(k^{2}+k^{2}z+z)\;,
\nonumber\\
\beta_\mu & \equiv  & \sqrt{1-\frac{4\hat{m}_{\mu}^2}{z}}\; ,
\end{eqnarray}
and  $\theta$ is the angle between the 
momenta of $K$ meson and $\mu^-$ in the dilepton 
centre of mass frame. 
The parameters $A,B,C,D,E,F,G$ are combinations of the Wilson
coefficients and the form factors, given by
\begin{eqnarray}
A & \equiv & 2C^{eff}_9\,f_{+}(z)-4C^{eff}_7\hat{m}_b \frac{f_{T}(z)}{1+k}\;,
\nonumber\\
B & \equiv  & 2C_{10}\, f_{+}(z)\;,
\nonumber\\
C & \equiv  & 2C_{10}\,\frac{1-k^2}{z}\Big[f_{0}(z)-f_{+}(z)\Big] \;,
\nonumber\\
D & \equiv &2R_S\frac{m_B(1-k^2)}{\hat{m}_b}f_0(z)  \;,
\nonumber\\
E & \equiv& 2R_P\frac{m_B(1-k^2)}{\hat{m}_b}f_0(z)  \;,
\nonumber\\
F & \equiv & -4C_T\frac{f_T(z)}{m_B(1+k)}\;,
\nonumber\\
G & \equiv  & 4C_{TE}\frac{f_T(z)}{m_B(1+k)}\; .
\end{eqnarray}

The kinematical variables in eq.~(\ref{double_drate}) are bounded as 
\begin{equation} 
-1\leq \cos\theta\leq 1,\;\;
4\hat{m}^2_{\mu}\leq z \leq(1-k)^2\; .
\label{kine-bounds}
\end{equation} 
The form factors $f_{+,\,0,\,T}$ can be calculated in the
light cone QCD approach. Their $z$ dependence 
is given by \cite{ali-00} 
\begin{eqnarray} 
f(z)=f(0)\,\exp(c_1z+c_2z^2+c_3z^3)\;, 
\end{eqnarray} 
where the parameters $f(0), c_1$, $c_2$ and $c_3$ for each form 
factor are given in Table~\ref{ff-table}.  
\begin{table}[h!] 
$$ 
\begin{array}{l c c c c} 
\hline 
    & \phantom{-}f(0)  &\phantom{-}c_1  & \phantom{-}c_2  & \phantom{-}c_3\\ \hline 
	f_{+}&\phantom{-}0.319^{+0.052}_{-0.041}  &\phantom{-}1.465  &\phantom{-}0.372 
&\phantom{-}0.782\\ 
f_0 &\phantom{-}0.319^{+0.052}_{-0.041}   &\phantom{-}0.633 
&\phantom{-}-0.095  &\phantom{-}0.591\\ 
f_T &\phantom{-}0.355^{+0.016}_{-0.055} 
&\phantom{-}1.478 & \phantom{-}0.373   &\phantom{-}0.700 \\ \hline 
\end{array} 
$$ 
\caption{Form factors for the $B \to K$ transition \cite{ali-00}.
\label{ff-table}} 
\end{table}

The FB asymmetry arises from the $\cos\theta$ term in 
the last two lines of eq.~(\ref{double_drate}).  
We get
\begin{equation}
A_{FB}(z)=\frac{2\Gamma_0 \, \beta_\mu \, \phi \, N(z)}{d\Gamma / dz}\;,
\label{afb}
\end{equation}
where
\begin{equation}
\Gamma_0=\frac{G_F^2\alpha^2}{2^{12}\pi^5}|V_{tb}V^*_{ts}|^2m_B^5\;,
\end{equation}
\begin{eqnarray}
N(z) &=& -4m_\mu(1-k^2) {\rm Re}(BG^*)-\frac{\hat{m}_\mu}{m_B}{\rm Re}(AD^*)-4z\hat{m}_\mu m_B{\rm Re}(CG^*)\nonumber\\
&& - \frac{z}{4}{\rm Re}(DF^*)-2z{\rm Re}(EG^*) \; ,
\end{eqnarray}
\begin{eqnarray}
\frac{d\Gamma}{dz} &  = & \Gamma_0\, \phi^{1/2}  
\times
\Bigg[\phi\left(1-\frac{1}{3}\beta_\mu^2\right)\,(|A|^2+|B|^2)\, +\, 4\,\hat{m}_{\mu}^2\,|B|^2\,(2+2k^2-z)\,+\, 4\,\hat{m}_{\mu}^2\,z\,|C|^2\, 
\nonumber \\
&& +\,8\,\hat{m}_{\mu}^2\,(1-k^2)\,{\rm Re} (BC^*)\,+\,8\hat{m}_{\mu}\,m_B\phi \,{\rm Re}(AF^*)\,+\,\frac{z}{m_B^2}\,(|E|^2+\beta_\mu^2\,|D|^2)
\nonumber \\
&& +\frac{4\hat{m}_\mu}{m_B}\,(1-k^2)\,{\rm Re}(BE^*)+\frac{4\hat{m}_\mu}{m_B}\, z\, {\rm Re}(CE^*)
\nonumber \\
&& +\, \frac{4}{3}\,\phi\,z\,m_B^2\, \left\{3|F|^2\,+\,2\,\beta_\mu^2\,(2|G|^2-|F|^2)\, \right\}
\Bigg]\;.
\end{eqnarray}
In our analysis we assume that there are no additional CP phases 
apart from the single Cabibbo-Kobayashi-Maskawa (CKM) phase. Under this assumption 
the new physics couplings are all real.

\section{$A_{FB}$ from new scalar/pseudoscalar operators }
\label{spnp}

If new physics is only in the form of scalar/pseudoscalar operators, then 
$\dfb$ is obtained by putting $C_T=C_{TE}=0$ in 
eq.~(\ref{matrix}). We get
\begin{equation}
\dfb=\frac{\beta_\mu\,\phi^{1/2}\,a_{SM,S}(z)\,R_S}{b_{SM}(z)\,+\,b_{SM,S}(z)R_P\,+\,b_{S}(z)(R_S^2+R_P^2)}\;,
\end{equation}
where
\begin{eqnarray}
a_{SM,S}(z) &=&-\frac{4\hat{m}_\mu}{\hat{m}_b}\, (1-k^2)\,f_0(z)\,{\rm Re}(A)\;, \\
b_{SM}(z) &=& \phi\left(1-\frac{1}{3}\beta_\mu^2\right)\,(|A|^2+|B|^2)\, +\, 4\,\hat{m}_{\mu}^2\,|B|^2\,(2+2k^2-z)
\nonumber\\
&& + \, 4\,\hat{m}_{\mu}^2\,z\,|C|^2\,+\,8\,\hat{m}_{\mu}^2\,(1-k^2)\,{\rm Re} (BC^*)\;, 
\label{bsm}\\
b_{SM,S}(z) &=&\,\frac{16\hat{m}_\mu}{\hat{m}_b}(1-k^2)^2\,C_{10}\,f_{0}^2(z)\;, 
\\
b_{S}(z)&=&\frac{4\,z}{\hat{m}_b^2} (1-k^2)^2\,f_{0}^2(z)\;.
\end{eqnarray}
Therefore in order to estimate $\dfb$ we need to know the 
scalar/pseudoscalar couplings $R_S$ and $R_P$.

\begin{figure} 
\centering
\includegraphics[width=0.6\textwidth,angle=270]{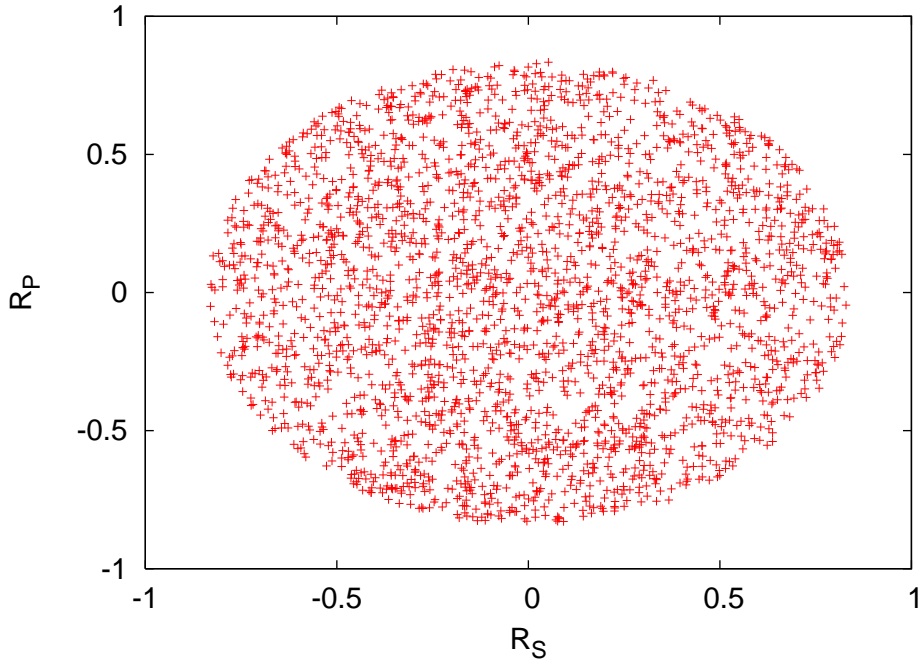} 
\caption{$R_S-R_P$ parameter space allowed by 
the present upper bound on the branching ratio of $\bsll$ 
\label{rsrp_fig}}
\centering 
\end{figure}

We constrain $R_S$ and $R_P$ through the decay
$\bsll$. The branching ratio of $\bsll$ due to $L_{SM}+L_{SP}$ is given by \cite{alok-02}
\begin{equation}
B(\bsll) = \frac{G_F^2\,\alpha^2\,m_{B_s}^3\,\tau_{B_s}}{64\pi^3}|V_{tb}V_{ts}^*|^2\,f_{B_s}^2
           \times \left[R_S^2+\left(R_P\,+\,2 \hat{m}_\mu\,C_{10}\right)^2 
\right] \; .
\end{equation}
The present upper bound on $B(\bsll)$ is \cite{cdf-07}
\begin{equation}
B(B_s \rightarrow \mu^+ \mu^-) < 0.58 \times10^{-7}
\quad  (95\% ~{\rm C.L.}) \; ,
\label{mumu-lim}
\end{equation}
which is still more than an order of magnitude 
away from its SM prediction.
Therefore we will neglect the SM contribution while obtaining constraints 
on the $R_S-R_P$ parameter space. 
The allowed values of $R_S$ and $R_P$ at $2\sigma$ 
are shown in Fig.~\ref{rsrp_fig}. The input values of parameters,
used throughout this paper, are given in Table~\ref{tab:inputs}.

\begin{table} 
\begin{center} 
\begin{displaymath} 
\begin{tabular}{lll} 
\hline \spp $G_F = 1.166 \times 10^{-5} \; \gev^{-2}$ & \phantom{spc} &
     $m_{B_s}=5.366 \; \gev$ \\ 
\spp $\alpha = 1.0/129.0$ & &
     $ m_B=5.279 \; \gev$  \\ 
\spp $\alpha_s(m_b)=0.220$ \cite{beneke-99} & & 
     $V_{tb}= 1.0 $  \\ 
\spp $\tau_{B_s} = 1.45 \times 10^{-12}\; s$  & & 
     $V_{ts}= (40.6 \pm 2.7) \times 10^{-3}$ \\ 
\spp $m_{\mu}=0.105 \;\gev$ & &
     $\left|V_{tb}V_{ts}^*/V_{cb}\right|=0.967\pm0.009$ \cite{charles} \\ 
\spp $m_K= 0.497 \; \gev$ & &
     $m_c/m_b=0.29\;$ \cite{ali-02} \\ 
\spp $m_b=4.80\; \gev $ \cite{ali-02} & & 
     $B(B \to X_c \ell \nu)=0.1061\pm0.0016\pm0.0006$ \cite{Aubert:2004aw}\\ 
     \hline 
\end{tabular} 
\end{displaymath} 
\caption{Numerical inputs used in our 
  analysis. Unless explicitly specified, they are taken from the 
  Review of Particle Physics~\cite{Yao:2006px}.\label{tab:inputs}} 
\end{center} 
\end{table} 

\begin{figure}
\centering
\includegraphics[width=0.6\textwidth,angle=270]{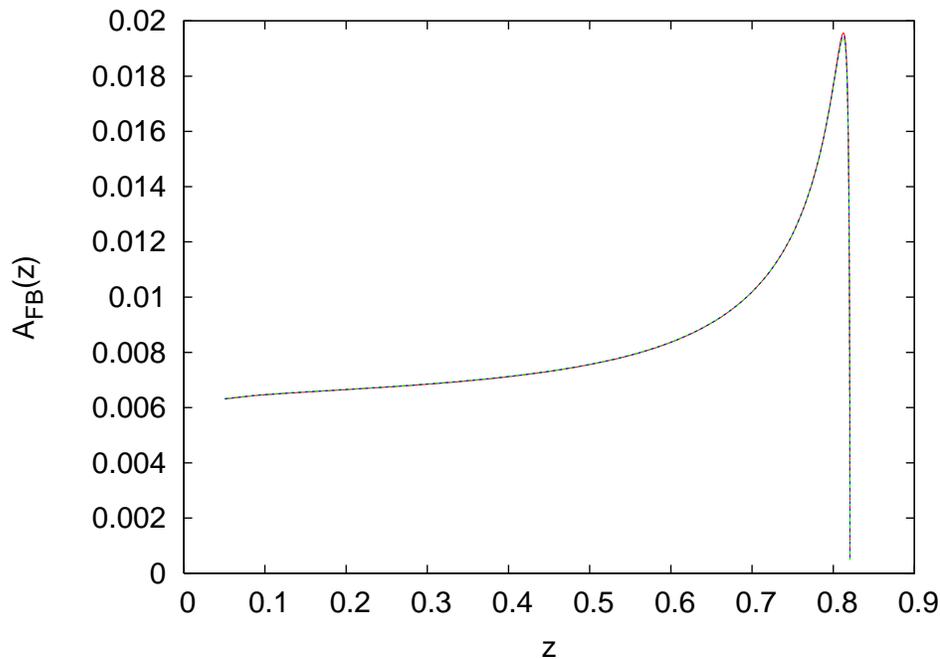} 
\caption{The forward-backward asymmetry $A_{FB}(z=q^2/m_B^2)$ for the new 
physics only in the form of scalar/pseudoscalar operators. 
The plot corresponds to $R_P=0$ and $R_S=-0.84$. 
The red (solid) curve corresponds to the central
values of the the form factors given in Table~\ref{ff-table} 
whereas the green (dashed) and blue (dotted) curves correspond to
their values at $+2\sigma$ and $-2\sigma$ respectively. 
In this scenario, all the curves overlap, indicating that the dependence
on form factors is negligibly small.
\label{spnp_fig}} 
\centering 
\end{figure}

The maximum value of $\dfb$ is obtained for $R_P=0$ and $R_S=\pm0.84$. 
At these parameter values, $\dfb$ is shown in Fig.~\ref{spnp_fig} 
for the central and $\pm 2\sigma$ values of the form factors. 
As can be observed, the errors in the form factors have almost
no impact on the value of $\dfb$ obtained.
The peak value of $\dfb$ is observed to be $\approx 2\%$,
whereas in most of the $z$ range, $\dfb < 1\%$.
Measurement of $\dfb$ in the presence of only scalar/pseudoscalar
operators will therefore be very challenging.

\section{$A_{FB}$ from new tensor operators }
\label{tpnp}

If new physics is only in the form of tensor operators then $\dfb$ is obtained by putting $R_S=R_P=0$ in 
eq.~(\ref{matrix}). We get
\begin{equation}
\dfb=\frac{\beta_\mu\,\phi^{1/2}\,a_{SM,T}(z)\,C_{TE}}{b_{SM}(z)\,+\,b_{SM,T}(z)C_T\,+\,b_{T}(z)(C_T+4C_{TE}^2)}\;,
\end{equation}
where
\begin{eqnarray}
a_{SM,T}(z)&=&-64\,\hat{m}_\mu \, (1-k)\,C_{10}\,f_T(z)\,f_0(z)\;,
\\
b_{SM,T}(z)&=&-\,\frac{32\,\hat{m}_\mu\,\phi\,{\rm Re}(A)\,f_T(z)}{1+k}\;,
\\
b_{T}(z)&=&\frac{64\,\phi\,z\,f_T^2(z)}{3\,(1+k)^2}\;,
\end{eqnarray}
and $b_{SM}(z)$ is given already in eq.~(\ref{bsm}).

In order to estimate $\dfb$, we need to know the tensor couplings
$C_T$ and $C_{TE}$.
In \cite{alok-03}, it was shown that the the most stringent bound 
on tensor couplings comes from the data on the branching ratio of 
the inclusive decay $\incl$. 
The branching ratio of $B \to X_s(p_s) \mu^+(p_{\mu^+}) \mu^-(p_{\mu^-})$  
is given by \cite{fukae}
\begin{equation}
B(B \to X_s l^+ l^-) = B_{0}
\Bigg[I_{SM}+(C_{T}^2+4C_{TE}^{2})I_T \Bigg]\,,
\label{incl_onlyt}
\end{equation}
where
\begin{eqnarray}
I_{SM} & = & \int dz \, \Bigg[\frac{8u(z)}{z}\left\{1-z^2+\frac{1}{3}u(z)^2\right\}C_{7}^{\rm eff}\nonumber \\
&& -2\,u(z) \left\{z^2+\frac{1}{3}u(z)^2-1\right\}({C_9^{\rm eff}}^2+C_{10}^2)  \nonumber\\
&& -16\,u(z)\,(z-1)\,C_{9}^{\rm eff}\,C_7^{\rm eff} \Bigg]\;, \\
I_T&=&16\int dz \, u(z)\Bigg[\frac{-2}{3}u(z)^{2}-2z+2\Bigg]\;,
\\
u(z)&=&(1-z)\;.
\end{eqnarray}
Here $z \equiv q^2/m_b^2 =(p_{\mu^+}+p_{\mu^-})^2/m_b^2 =
(p_b-p_s)^2/m_b^2$. 
The limits of integration for $z$ are now
\begin{equation} 
z_{min}=4m_{\mu}^2/{m_b}^2 \; , \quad z_{max}=(1-\frac{m_s}{m_b})^2 \; ,
\end{equation}
as opposed to the ones given in eq.~(\ref{kine-bounds}) for 
the exclusive decay.
The normalization factor $B_0$ is given by
\begin{eqnarray}
B_0 = B(B\to X_c e \nu) \frac{3 \alpha^2 }{16 \pi^2 }
             \frac{ |V_{ts}^*V_{tb}|^2 }{|V_{cb}|^2 } 
            \frac{1}{f(\hat{m_c}) \kappa(\hat{m_c})}\; ,
\end{eqnarray}
where the phase space factor $f(\hat{m_c}={m_c \over m_b})$, 
and the $O(\alpha_s)$
QCD correction factor $\kappa(\hat{m_c})$ 
of $b \rightarrow c e \nu$ are given by \cite{Kim}
\begin{eqnarray}
f(\hat{m_c}) &=& 1 - 8 \hat{m_c}^2 + 8 
        \hat{m_c}^6 - \hat{m_c}^8 - 24 \hat{m_c}^4 \ln \hat{m_c} \;,  \\
\kappa(\hat{m_c}) &=& 1 - \frac{2 \alpha_s(m_b)}{3 \pi} 
           \left[(\pi^2-\frac{31}{4})(1-\hat{m_c})^2 + \frac{3}{2} \right]\; . 
\end{eqnarray}
Eq.~(\ref{incl_onlyt}) can be written as
\begin{equation}
B(\incl)=B_{SM}(\incl)+B_T(\incl)\;,
\label{incltot}
\end{equation}
where
\begin{eqnarray}
B_{SM}(\incl) & = & B_0\,I_{SM}\; , \\
B_T(\incl) & = & B_0\,I_T\,(C_T^2+4C_{TE}^2)\;.
\end{eqnarray}

\begin{figure} 
\centering
\includegraphics[width=0.6\textwidth,angle=270]{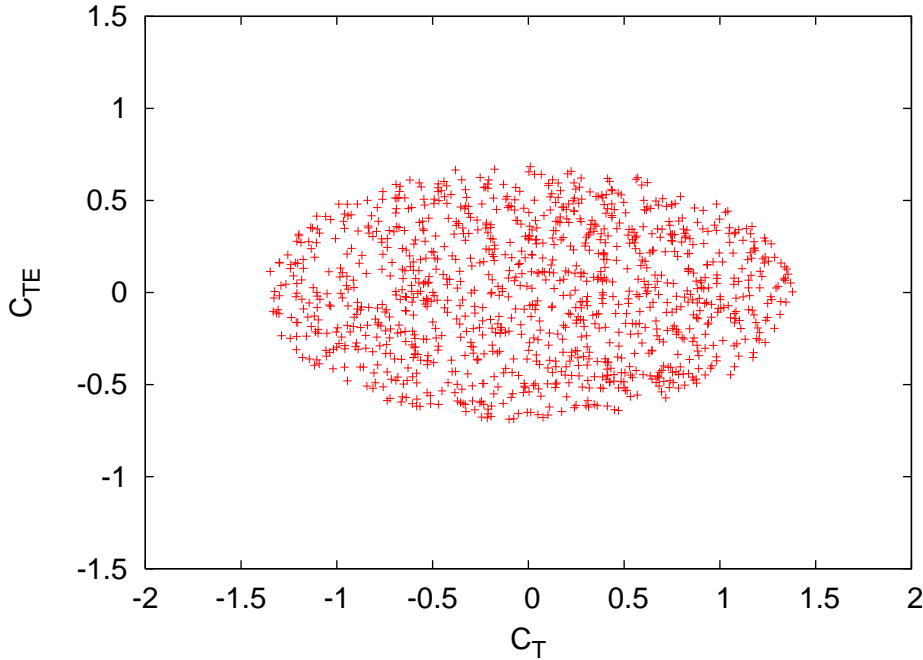} 
\caption{$(C_T,C_{TE})$ parameter space at $2 \sigma$ allowed by the measurement of branching ratio of 
$\incl$
\label{ctcte_fig} } 
\centering 
\end{figure} 
\begin{figure}
\centering
\includegraphics[width=0.6\textwidth,angle=270]{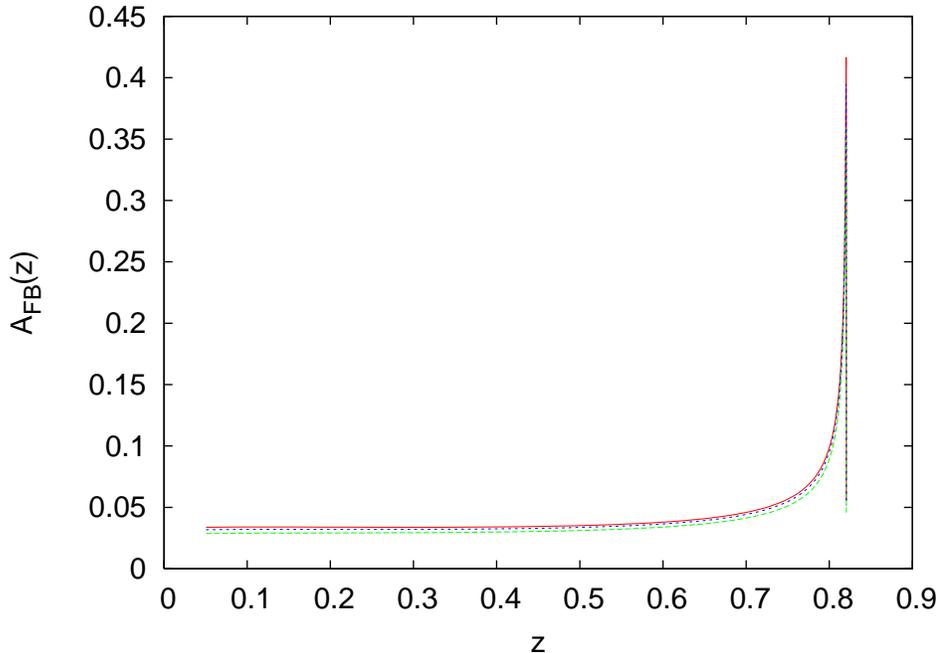} 
\caption{The forward-backward asymmetry $A_{FB}(z=q^2/m_B^2)$ for 
the new physics only in the form of tensor operators. The plot corresponds to $C_T=0$ and $C_{TE}=+0.69$.
The red (solid) curve corresponds to the central
values of the the form factors given in Table~\ref{ff-table} 
whereas the green (dashed) and blue (dotted) curves correspond to
their values at $+2\sigma$ and $-2\sigma$ respectively. 
The dependence on the form factors is clearly extremely small.
\label{tpnp_fig}}
\centering 
\end{figure} 

The present world average for $B(\incl)$ is \cite{hfag}
\begin{equation}
B_{\rm Exp}(\incl)_{q^2>0.04\,{\rm GeV^2}}=(4.3^{+1.3}_{-1.2})\times 10^{-6}\;.
\label{inclexp}
\end{equation}
We keep the same invariant mass cut, $q^2 > 0.04$ GeV$^2$, 
in order to enable comparison with the experimental data. With this
range of $q^2$,
the SM branching ratio for $\incl$ in NNLO is \cite{ali-02}
\begin{equation}
B_{SM}(\incl)_{q^2>0.04\,{\rm GeV^2}}=(4.15\pm0.71)\times 10^{-6}\; ,
\label{inclsm}
\end{equation}
whereas $B_0 I_T = (1.47\pm0.22)\times 10^{-6}$.
Using equations (\ref{incltot}), (\ref{inclexp}) and (\ref{inclsm}), we get
\begin{equation}
C_T^2 + 4C_{TE}^2=0.10\pm1.01 \; .
\end{equation}
The allowed parameter space for $C_T, C_{TE}$ at $2\sigma$ is shown in Fig.~\ref{ctcte_fig}.

The maximum value of $\dfb$ is obtained for $C_T=0$ and $C_{TE}=\pm0.69$. 
For these parameter values, $\dfb$ is shown in  
Fig.~\ref{tpnp_fig} for the central and $\pm 2\sigma$ values of the 
form factors.
In most of the $z$ range, $\dfb \lesssim 3\%$, however 
its peak value at the high-$q^2$ end point is $\sim 40\%$. 
Thus there can be a large deviation from the SM prediction 
in the high-$q^2$ region.

\section{$A_{FB}$ from the combination of scalar/pseudoscalar 
and tensor operators}
\label{sptp}

\begin{figure} 
\centering
\includegraphics[width=0.6\textwidth,angle=270]{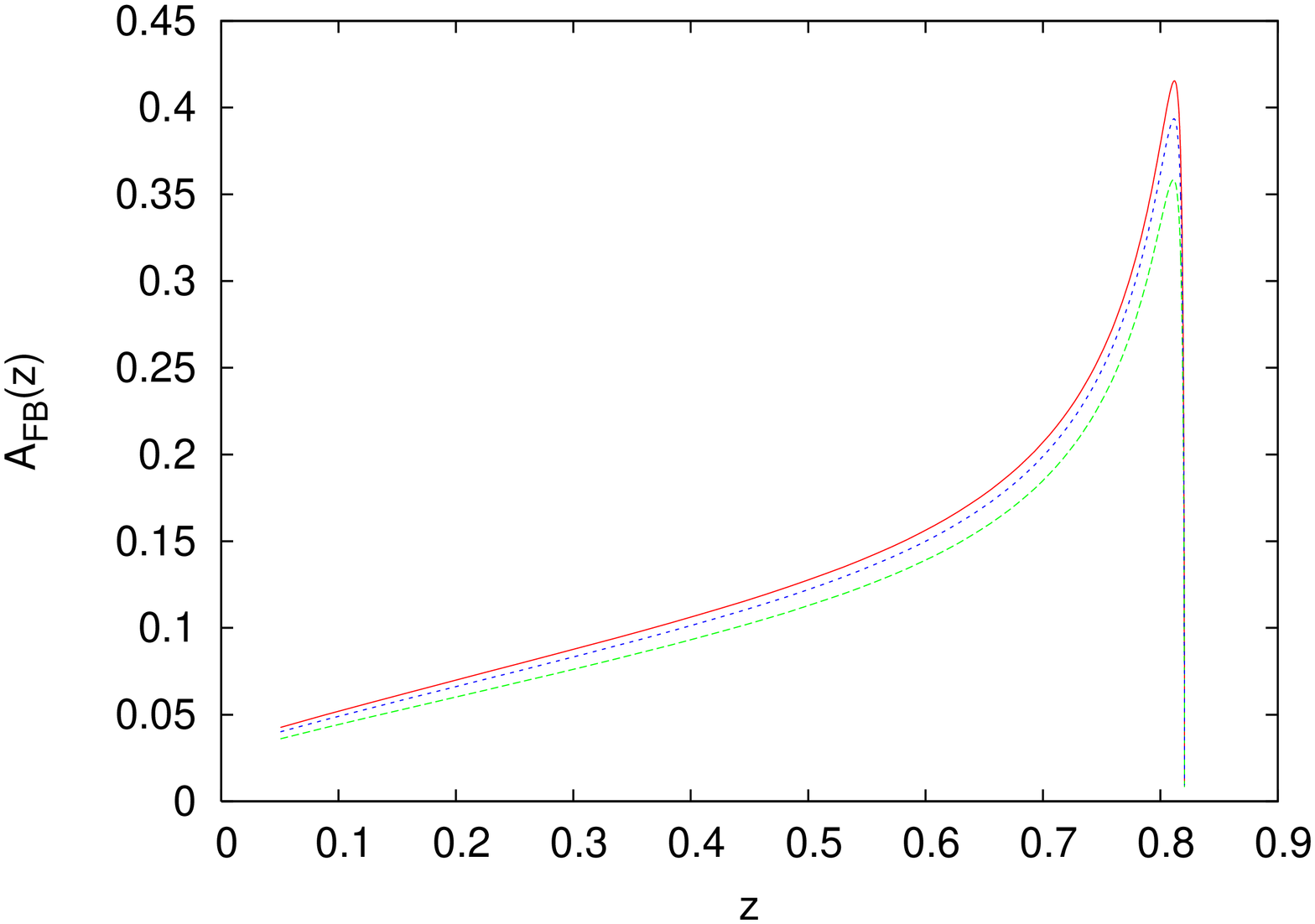} 
\caption{The forward-backward asymmetry $A_{FB}(z=q^2/m_B^2)$ for new physics 
when both scalar/pseudoscalar as well as tensor operators are present. 
The plot corresponds to $R_S=C_T=0$ and $R_P=-0.84,\,C_{TE}=+0.69$.
The red (solid) curve corresponds to the central
values of the the form factors given in Table~\ref{ff-table} 
whereas the green (dashed) and blue (dotted) curves correspond to
their values at $+2\sigma$ and $-2\sigma$ respectively. 
\label{sptp_fig}} \centering 
\end{figure} 

We now consider the scenario where new physics in the form of both 
scalar/pseudoscalar and tensor operators are present.
In this case the expression for $\dfb$ is given by eq.~(\ref{matrix}). 
Maximum values of $\dfb$ as obtained for 
$R_S=C_T=0$ and $R_P=-0.84,\,C_{TE}=0.69$, which are 
shown in Fig.~\ref{sptp_fig}.
The peak value of $\dfb$ is $\sim 40\%$ 
at $2\sigma$ and is obtained at the high-$q^2$ end point. 
Thus, there can be large FB asymmetry in the high 
$q^2$ region.
Another reason to concentrate on the high-$q^2$ region is that
theoretical predictions of the decay rate $\bkll$ are more 
robust there, owing to the non-interference of charmed resonances.

Let ${\cal R}$ be the high-$q^2$ region, with $q_0< q^2 < q^2_{\rm max}$,
where $q^2_{\rm max}$ is the endpoint. 
The restriction to high-$q^2$ would decrease the number of
events selected, however since the average $A_{FB}$ in this region,
$\langle A_{FB}^{\cal R} \rangle$, is larger, it can still be observed.
The number of events
of $\bkll$ required to determine this asymmetry to $n\sigma$ is 
\begin{equation}
N_{\bkll} \gtrsim \frac{n^2}{\langle A_{FB}^{\cal R} \rangle^2 f^{\cal R}} \;,
\label{n-events}
\end{equation}
where $f^{\cal R}$ is the fraction of total number of $\bkll$ events
that lie in the region ${\cal R}$.
When ${\cal R}$ corresponds to the whole $q^2$ range available, then the
expression reduces to $N_{\bkll} \gtrsim n^2/\langle A_{FB} \rangle^2$,
as expected.

Taking ${\cal R}$ to be the region $q^2 > 15 $ GeV$^2$ and the values of 
parameters as shown in Fig.~\ref{sptp_fig}, we find that about 600 
total $\bkll$ events are required to observe FB asymmetry at $2\sigma$.
For $q^2 > 19$ GeV$^2$, the required number of events for $2\sigma$ detection 
of $A_{FB}$ is about 1600. 
These numbers are easily obtainable at a Super-$B$ factory as well
as at the LHC, so the structure of the $A_{FB}(q^2)$ peak can be
studied at these experiments.

\section{Conclusions}
\label{concl}

In the standard model, the forward-backward asymmetry $A_{FB}$
of muons in $B \rightarrow K \mu^+ \mu^-$ is negligible. 
New physics in the form of vector/axial vector operators also
cannot contribute to $A_{FB}$. 
However, new physics in the form of scalar/pseudoscalar or tensor
operators can enhance $A_{FB}$ to per cent level or more,
thus bringing it within the reach of the LHC or a Super-$B$ factory.
In this paper, we concentrate on the magnitude as well as
$q^2$ dependence of $A_{FB}$ with these kinds of new physics.

We find that if new physics is in the form of 
scalar/pseudoscalar operators only, 
then the peak value of $A_{FB}(q^2)$ can only be $ \lesssim 2\%$,
and hence rather challenging to detect.
However if new physics is only in the form of tensor operators then 
the peak value of $A_{FB}(q^2)$ can be as high as $40\%$. 
Such a high enhancement is obtained only near the 
high-$q^2$ end point, i.e. for $q^2 > 19$ GeV$^2$, below which
$A_{FB}(q^2) \lesssim 5\%$.
In the presence of both scalar/pseudoscalar and tensor operators, 
the interference terms between them can boost $A_{FB}(q^2)$ to more than
$15\%$ for the whole region $q^2 > 15$ GeV$^2$.

The measurement of the distribution of $A_{FB}$ as a function 
of $q^2$ can not only reveal new physics, but also
indicate its possible Lorentz structure.
A large enhancement in $A_{FB}$ by itself would confirm the presence
of new physics tensor operators. If the enhancement is only at 
large $q^2$ values, the scalar/pseudoscalar new physics operators
probably play no major role. On the other hand, if the enhancement
as a function of $q^2$ is significant at low $q^2$ and increases gradually
with increasing $q^2$, the presence of scalar/pseudoscalar new physics
operators would be indicated.

The high-$q^2$ region in the $A_{FB}(q^2)$ distribution is 
theoretically clean since the charmed resonances in the
intermediate $q^2$ region do not interfere here.
This region also happens to be highly sensitive to new physics,
especially in the form of tensor operators, as we have
shown here.
Exploration of this region in the upcoming experiments
is therefore of crucial importance.

\acknowledgments

A.K. Alok would like to acknowledge J. Matias, E. Laenen 
and G. Hiller for useful discussions. 
He would also like to thanks Nicola Serra and  Fabian Jansen for valuable suggestions.

\end{document}